\documentclass[a4paper,11pt]{article}

\usepackage{jheppub,amsmath, amssymb,amsthm,dsfont,graphicx,mathtools,multirow,rotating,placeins,verbatim} 

\usepackage[normalem]{ulem}

\usepackage{tikz}
\usetikzlibrary{arrows.meta, positioning}

\usepackage{hyperref}
\usepackage{comment}
\usepackage{color}

\usepackage{setspace}
\usepackage{cancel}
\usepackage{graphicx}
\usepackage{enumerate} 

\tikzset{myarrow/.style={->, shorten >=5pt, shorten <=5pt, >=Latex,thick}}

\def\be{\begin{equation}}
\def\ee{\end{equation}}
\def\ba{\begin{eqnarray}}
\def\ea{\end{eqnarray}}
\def\bal{\begin{equation} \begin{aligned}}
\def\eal{\end{aligned}\end{equation} }

\def\bi{\begin{itemize}}
\def\ei{\end{itemize}}

\def\w{\omega}

\def\nh{\hat{n}}

\def\Lie{\mathcal{L}}
\def\w{\omega}

\newcommand{\ov}[2]{\overset{\scriptscriptstyle #1}{#2}\vphantom{#1}}
\def\t{\tau}

\def\T{\mathcal{T}}

\def\harm{\text{harm}}
\def\rad{\text{BS}}

\def\G{\mathcal{G}}

\def\scri{\mathcal{I}}
\def\massive{\text{massive}}
\def\massless{\text{massless}}
\def\total{\text{total}}
\def\grav{\text{grav}}

\def\czero{\mathring{c}}
\def\sigmazero{\mathring{\sigma}}
\def\div{\text{div}}
\def\rad{\text{rad}}

\def\E{\mathcal{E}}
\def\inn{\text{in}}
\def\out{\text{out}}

\def\B{\mathcal{B}}
\def\tree{\text{tree}}
\def\Rt{\tilde{R}}
\def\Rt{\tilde{r}}

\title{Log translation invariance of  log soft gravitational radiation}

\author[a]{Gianni Boschetti}
\author[b]{Miguel Campiglia}

\affiliation[a]{Instituto de Física, Facultad  de  Ingeniería, Universidad  de  la  Rep\'ublica 
Julio Herrera y Reissig 565,  Montevideo,  Uruguay}

\affiliation[b]{Instituto de Física, Facultad  de  Ciencias, Universidad  de  la  Rep\'ublica 
Ig\'ua  4225,  Montevideo,  Uruguay}

\emailAdd{gboschetti@fing.edu.uy}
\emailAdd{miguel.campiglia@fcien.edu.uy}

\abstract{The gravitational radiation emitted during a classical scattering process is known to exhibit two universal logarithmic terms in its soft frequency expansion. We show that these terms   can be written in a way that makes the action of \emph{logarithmic translations} manifest.  Invariance under log translations naturally explains a puzzling cancellation in the contribution from outgoing massless particles and leads to  a recurrence relation for expected higher-order universal log soft terms.}

\begin{document}
\maketitle
\flushbottom

\tableofcontents

\section{Introduction}

In a series of remarkable papers \cite{laddhasen1,laddhasen,sahoosen,proofdeq4,sahoosubsub,rewritten} Laddha, Saha, Sahoo, and Sen identified universal logarithmic terms in the low-frequency expansion of gravitational radiation, along with the associated early/late-time tails; see~\cite{senreview} for a review. The  universal nature of the first log term can be understood in terms of  superrotations  \cite{Agrawal:2023zea,chipum,chipum2}, within the broader setting of an ``infrared triangle'' that connects  asymptotic symmetries with soft theorems and memory effects \cite{IRlectures}. 

In this paper, we  revisit these \emph{classical soft theorems} from a different perspective. Our motivation lies in  the seemingly ``miraculous'' cancellation of the outgoing massless contributions to the log soft factors \cite{sahoosen,rewritten}.  This cancellation has so far been observed in the first two known universal log terms but is believed to extend to conjectured higher-order terms \cite{sahoosubsub,spectra,rewritten,senreview}.  Here we provide a simple explanation of this cancellation  by  making  explicit a  gauge redundancy in the asymptotic description of the gravitational field known as logarithmic translations \cite{Bergmann:1961zz,BS,aalog}.\footnote{These are distinct from log supertranslations, recently proposed  as part of the symmetry algebra of asymptotically flat gravity \cite{fuentelog} (see also \cite{virmanilog,kinjal}). Although we do not make use of this extension, our perspective is consistent with~\cite{fuentelog} in treating log translations as pure gauge.} In a nutshell, the contribution from outgoing massless particles can be removed altogether by a log translation. But since the  soft factors are invariant under such transformations, this contribution must vanish to begin with.

Logarithmic translations were first introduced in the context of spatial infinity, but they can also arise at timelike and even null infinity, as we discuss in the next section. Our analysis draws from Ashtekar’s geometric treatment of spatial infinity~\cite{hansen,heldvol,romano}, exported to the case of timelike infinity, following the spirit of the “asymptotic framework” of Compère, Gralla, and Wei~\cite{cgw} (see also~\cite{persides,cutler,Gen:1997az,clmassive,virmani,Figueroa-OFarrill:2021sxz}). The ideas presented here set the stage for a fully asymptotic derivation of the log soft theorems, an approach we develop in \cite{gianni2}.

The organization of the paper is as follows. In section \ref{loglogsec} we review the concept of log translations and introduce the notion of log deviation vector, which captures the leading deviation of asymptotic geodesics from straight lines. In section \ref{hypcoordsec} we present the description of timelike infinity, and discuss how log translations arise in that context. We further argue that there is  a single ``global'' log translation group that simultaneously acts on the asymptotic future and past. Given these preliminaries, in section \ref{logsoftsec} we show how the  two known log soft theorems can be  expressed in terms of the log deviation vector in a log translation invariant way.  In section   \ref{higherordersec} we explore the consequences of log translation invariance on the putative higher order log soft theorems. We conclude in section \ref{finalsec} with a brief discussion of our results and future perspectives. The paper is complemented with two appendices containing supporting details for the main discussion.

\section{Log translations and log deviation vector} \label{loglogsec}

Any notion of asymptotic flatness requires that the spacetime metric becomes flat as some distance parameter \( R \) goes to infinity. While the precise meaning of \( R \) varies with the setting---in particular, with the signature of the direction along which infinity is approached---the leading deviation from the flat metric typically scales as \( 1/R \),
\begin{equation} \label{geta1overR}
g_{\mu \nu} \stackrel{R \to \infty}{=} \eta_{\mu \nu} + O(1/R),
\end{equation}
where \( g_{\mu \nu} \) is the spacetime metric in appropriate asymptotic Cartesian coordinates, and \( \eta_{\mu \nu} \) the Minkowski metric.  Depending on the gauge condition being used, there may exist a residual gauge freedom affecting the \( 1/R \) term in \eqref{geta1overR} due to \emph{logarithmic translations} \cite{Bergmann:1961zz,aalog},
\begin{equation} \label{deflogvf}
\xi^\mu_L \stackrel{R \to \infty}{=} \log R\, L^\mu + O(R^0),
\end{equation}
where \( L^\mu \) are constant vectors.
  A well-studied instance where this occurs is in Beig–Schmidt gauge at spatial \cite{BS} and timelike \cite{virmani,cgw} infinity.  By contrast, radiative gauges  at null infinity---such as Bondi \cite{bondi,sachs} or Newman–Unti \cite{NU}---do not allow for log translations.  A second relevant example where log translations are frozen  is harmonic gauge (see appendix \ref{harmggeapp}). 

We will consider the standard situation in which the limit in \eqref{geta1overR} is approached along asymptotic geodesics,
\be \label{Xmulambda}
X^\mu(s) \stackrel{s \to \infty}{=} s V^{\mu} + \log  s  \, c^\mu + \cdots,
\ee
where  $V^{\mu}$ is the geodesic asymptotic direction and  $s$ an affine parameter that, to leading order, is proportional to $R$. The logarithmic term in \eqref{Xmulambda} arises from the $1/R$ piece in the metric \eqref{geta1overR}, which leads to non-trivial $O(1/R^2)$ Christoffel symbols \cite{sahoosen}.\footnote{In the case of asymptotic null directions, we further require that $V^\mu V^\nu \partial_u g_{\mu \nu}=O(1/R^2)$, where $u$ is the associated asymptotic retarded (or advanced) time. This condition is satisfied in  solutions of Einstein equations in harmonic coordinates of the type discussed in \cite{sahoosen} and reviewed in section \ref{EEtimeinfsec}.} Indeed, using the geodesic equation and \eqref{Xmulambda} one can show that 
\be \label{CmuitoGamma}
c^\mu \equiv \lim_{R \to \infty}  \Gamma^\mu_{\nu \rho} X^\nu X^\rho.
\ee

Under logarithmic translations \eqref{deflogvf}, this \emph{log deviation vector}  transforms as
\begin{equation} \label{delLCmu}
\delta_L c^\mu = -L^\mu,
\end{equation}
as can be  inferred directly from  \eqref{Xmulambda}, or through a short calculation using the transformation properties of the Christoffel symbols.\footnote{We follow conventions such that the action of a vector field $\xi^\mu$ on the coordinates and the metric is given by $\delta_\xi X^\mu = - \xi^\mu$ and $\delta_\xi g_{\mu \nu} = \Lie_\xi g_{\mu \nu}$, respectively. The change of the $O(1/R)$ metric perturbation under log translations contains a piece from the $O(R^0)$ vector field in \eqref{deflogvf}, whose precise form depends on the gauge-fixing condition (see appendix~\ref{timeinfapp}). This dependence, however, drops out in the combination of Christoffel symbols given in \eqref{CmuitoGamma}.}

In general, the log deviation vector obtained in the limit \eqref{CmuitoGamma} depends on the asymptotic direction \( V^\mu \). A notable exception occurs for null rays, where \( c^\mu \) turns out to be independent of the null direction \cite{sahoosen}, see section~\ref{nullcmusec}.\footnote{A similar  contrast between null and spatial directions lies at the heart of the analysis  in \cite{hansen}.}  In this case, it can be gauged away by a logarithmic translation---precisely what is achieved in radiative coordinates. These considerations will play a key role in elucidating structural properties of the log soft theorem.

\section{Timelike infinity and log frames} \label{hypcoordsec} 

In order to apply the ideas of the previous section to a scattering setting, we  consider the case in which infinity is approached along timelike directions.\footnote{By an appropriate limiting procedure, this will also allow us to describe asymptotic null geodesics, and hence massless particles.} In this limit, the leading structure of the gravitational field is captured by a scalar function on the unit hyperboloid that plays the role of a  gravitational potential. This potential encodes the same information as the log deviation vector but is simpler to determine from Einstein equations. We present two complementary methods for obtaining it: an ``asymptotic approach'', formulated independently of any specific gauge choice (aside from assuming the form~\eqref{geta1overR}), and a ``bulk approach'' that we implement in harmonic gauge.
The asymptotic approach naturally exposes the freedom under log translations, allowing one to select any desired log frame, such as those associated with radiative or harmonic coordinates.

At this stage, we will  essentially be recovering the leading term of the well-known coordinate transformation between harmonic and radiative gauges~\cite{blanchetnull,blanchetreview}. The novelty in our treatment arises when considering both the incoming and outgoing versions of these transformations, ultimately allowing us to identify \emph{global} log translations that simultaneously act at future and past infinities.
 This identification is a prerequisite for defining the action of log translations on soft factors,  discussed in  section \ref{logsoftsec}.

\subsection{Spacetime metric at timelike infinity} \label{timeinfsec}

For simplicity,  we  set the discussion in the context of future timelike infinity, but entirely analogous considerations apply to past time infinity. A comparison between the  future and past equations is presented in subsection \ref{pasttimesec}.

The role of the parameter \( R \) in \eqref{geta1overR} is  now played by the asymptotic proper time \( \tau \), while the asymptotic direction \( V^\mu \) is a unit timelike vector,
\begin{equation} 
X^\mu \stackrel{\tau \to \infty}{=} \tau V^\mu + \cdots, \quad V^\mu V_\mu = -1,
\end{equation}
where the dots indicate subleading terms and the norm condition is taken with respect to the Minkowski metric. The unit timelike hyperboloid is parametrized by coordinates \( x^a \) (see Eq. \eqref{xaitorotheta} below for an explicit expression)
\begin{equation} \label{Vmuitoxa}
V^\mu = V^\mu(x),
\end{equation}
and the associated hyperboloid metric is
\begin{equation}
h_{ab} = \partial_a V^\mu \partial_b V_\mu.
\end{equation}

We denote by $\gamma_{\mu \nu}(x)$ the tensor capturing the leading deviation from the flat metric in the large time limit,
 \begin{equation} \label{defgamma} 
g_{\mu \nu}(\t,x) \stackrel{\t \to \infty}{=} \eta_{\mu \nu} + \frac{1}{\t} \gamma_{\mu \nu}(x) + \cdots
\end{equation}
As discussed  in appendix \ref{timeinfapp}, this tensor   contains both ``pure gauge'' and ``non-gauge'' pieces. To disentangle them, it is convenient to decompose it in terms of 3d hyperboloid fields  
\begin{equation} \label{gamma3d}
\gamma_{\mu \nu} \leftrightarrow 
\left\{
\begin{array}{l}
\sigma \\
\gamma_a \\
\gamma_{ab}
\end{array}
\right.,
\end{equation}
where\footnote{In the notation of Appendix~\ref{hyperapp}, $\sigma=-\tfrac{1}{2} \gamma_\parallel$.} 
\begin{equation} \label{defsigma}
\sigma = -\tfrac{1}{2} V^\mu V^\nu \gamma_{\mu \nu}, \quad 
\gamma_a = \partial_a V^\mu V^\nu \gamma_{\mu \nu}, \quad 
\gamma_{ab} = \partial_a V^\mu \partial_b V^\nu \gamma_{\mu \nu}.
\end{equation}

The gauge invariant information contained in these fields can be obtained by evaluating the asymptotic electric and magnetic part of the Weyl curvature \cite{hansen}. Under typical asymptotic flatness conditions the latter vanishes, while the former is fully determined by $\sigma$. All remaining components in \eqref{gamma3d} are therefore either pure gauge or not independent.\footnote{Modulo a supertranslation Goldstone field \cite{phigoldstone}, which  plays no role for our analysis. See Appendix~\ref{timeinfapp} for a more detailed discussion of the material presented in this subsection.}

Summarizing, the non-trivial part of the metric perturbation at order $1/\t$ is the scalar \( \sigma \), which we refer to as the gravitational potential. Our interest in this quantity is that it  determines the log deviation vector according to,
\begin{equation} \label{citosigma}
c^\mu = D^a V^\mu \partial_a \sigma - V^\mu \sigma,
\end{equation}
as can be checked by direct computation from the definition \eqref{CmuitoGamma} for the case $R=\t$. 

Conversely, the potential can be recovered from the deviation vector via
\begin{equation} \label{sigmaitoc}
\sigma =  c^\mu V_\mu.
\end{equation}

Thus, $\sigma$ and $c^\mu$ encode the same information. Depending on the context, one or the other may be more convenient to work with.  Like the deviation vector,  $\sigma$ is not fully gauge invariant due to  logarithmic translations, which act as 
\begin{equation}\label{delLsigma}
\delta_L \sigma = - L^\mu V_\mu .
\end{equation}

\subsection{Null limit} \label{nullcmusec}
It will be of interest to consider the limit in which the timelike direction $V^\mu$ becomes null.   To be explicit, let us consider  coordinates  $x^a = (\rho, x^A)$ in \eqref{Vmuitoxa} defined  by
\begin{equation} \label{xaitorotheta}
V^\mu = (\sqrt{\rho^2+1}, \rho \, \nh), \quad V^\mu V_\mu = -1,
\end{equation}
where $\rho$ is a radial coordinate and \( x^A \) are  coordinates on the unit sphere associated with the unit 3-vector \( \nh \). In this parametrization, the hyperboloid line element takes the form
\be \label{habtime}
h_{ab} dx^a dx^b =\frac{d \rho^2}{\rho^2+1} + \rho^2 d \Omega^2
\ee
where $d \Omega^2$ is the unit sphere metric.

The null limit of $V^\mu$ is  achieved by taking $\rho \to \infty$ in \eqref{xaitorotheta}. To leading order one gets
\be \label{limrhoinfVmu}
V^\mu \stackrel{\rho \to \infty}{=} \rho \, n^\mu + O(1/\rho)
\ee
where 
\be \label{defnmu}
n^\mu =(1,\nh)
\ee
is the null direction associated to the unit 3-vector \( \nh \). The divergent $\rho$ factor multiplying the null direction \eqref{limrhoinfVmu}  indicates the  need to simultaneously rescale the affine parameter to properly describe the corresponding  null trajectory.   Alternatively, when  looking at a particle's momentum, \eqref{limrhoinfVmu} should be multiplied by a $m \to 0$ mass such that the energy $E= m \rho$ is kept constant. 

From a   3d hyperboloid perspective, the limit $\rho \to \infty$  brings us  to its asymptotic boundary. As we shall review in the next section,  Einstein equations imply that, in this limit,  $\sigma$ approaches  \cite{cgw}
\be
\sigma \stackrel{\rho \to \infty}{=}   \rho \,  n_\mu  c^\mu_\scri  + \cdots, 
\ee
where $c^\mu_\scri $ is  a constant (i.e. independent of $\nh$) vector.  It then follows from \eqref{citosigma} that this vector represents the large  $\rho$ limit of $c^\mu$,
\be
c^\mu_\scri \equiv \lim_{\rho \to \infty} c^\mu .
\ee

In fact, this vector coincides with the log deviation vector of null trajectories. This is because the logarithmic coefficient in the asymptotic geodesics \eqref{Xmulambda} is unaffected by rescalings of the affine parameter. The independence of \( c^\mu_\scri \) from \( \nh \) confirms the result quoted at the end of section \ref{loglogsec}, namely that for asymptotically null trajectories, the log deviation vector is independent of the null direction.

\subsection{Solving Einstein equations} \label{EEtimeinfsec}
\begin{figure}[ht]
  \centering
\begin{tikzpicture}[
  node distance=3cm and 4cm,
  >=Stealth,
  every node/.style={align=left},
  box/.style={draw, minimum width=3.5cm, minimum height=1cm, anchor=center},
  smallbox/.style={draw, minimum width=2.5cm, minimum height=0.8cm, anchor=center}]

\node[draw,  rounded corners, inner sep=7pt, thick] (eeqs) {Einstein eqs};
\node[draw,  rounded corners, inner sep=5pt, right=of eeqs, thick,fill=gray!15] (fullsol) {full metric};
\node[draw,  rounded corners, inner sep=5pt, below=of eeqs, thick,fill=gray!15] (asym) {Einstein eqs at \( i^+ \)};
\node[draw,  rounded corners, inner sep=7pt, thick, fill=gray!50,below=of fullsol] (asymsol) { metric at \( i^+ \)};

\draw[myarrow] (eeqs) -- node[above] {solve} (fullsol);
\draw[myarrow] (eeqs) -- node[left] {\( \tau \to \infty \)} (asym);
\draw[myarrow] (fullsol) -- node[right] {\( \tau \to \infty \)} (asymsol);
\draw[myarrow] (asym.east) -- node[above] {solve} (asymsol.west |- asym.east);

\node at (2,-2.7) {\small{\emph{asymptotic  path}}};
\node at (4.5,-1) {\small{\emph{bulk  path}}};


\end{tikzpicture}

  \caption{Two ways to find the asymptotic metric at timelike infinity $i^+$.}
  \label{twopaths}
\end{figure}
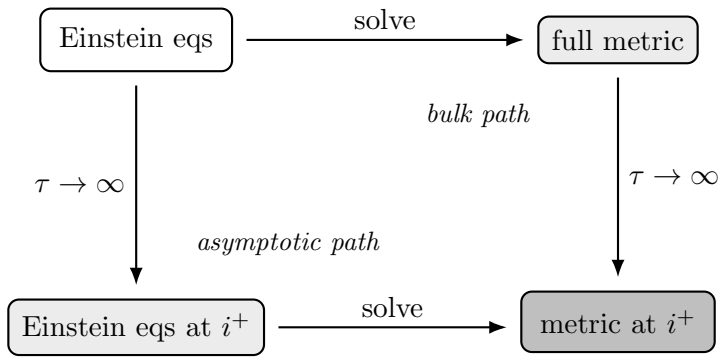

Our presentation so far  has been  mostly kinematical. We now  discuss how Einstein equations determine the gravitational potential (and more generally,  the spacetime metric) at timelike infinity. As illustrated in \autoref{twopaths}, there are two paths one may follow, depending on whether the large time limit is taken before or after solving Einstein equations.

Below, we discuss these two approaches, focusing on the gravitational potential. As we will see, the two methods generally yield potentials that differ by a log translation. The asymptotic approach will be presented in a way that is agnostic to gauge-fixing conditions. For the bulk approach, however, a gauge-fixing condition is necessary. We will consider the case of harmonic gauge, as used in standard perturbative treatments. This will allow us to characterize  the log translation  gauge-fixing condition implicit in the analysis of~\cite{laddhasen,sahoosen}, and eventually to relax it in section~\ref{logsoftsec}.

\subsubsection*{Asymptotic approach} \label{BSgaugeprelsec}

To leading order in the $\t \to \infty$ limit,  Einstein equations, applied to \eqref{defgamma} leads to (see e.g. \cite{cgw,chipum2,gianni2})
\be \label{eqsigma}
(D^2-3) \sigma = 4 \pi G \rho_\massive,
\ee
where $D^2$ is the hyperboloid Laplacian and 
\be \label{rhomassive}
\rho_\massive(x)=\sum_{\substack{i \\ m_i \neq 0}} m_i \delta(x,x_i)
\ee
is the energy density of massive particles at future timelike infinity. From the asymptotic perspective,  $\sigma$ is to be determined by solving \eqref{eqsigma}. Let $\G(x,x')$ be a  Green's function  for this equation, i.e. 
\be \label{d2minus3G}
(D^2-3) \G(x,x') =  \delta(x,x').
\ee
If we demand that $\G(x,x')$  depends only on the geodesic distance $d(x,x')$ between $x$ and $x'$, one arrives at  a one-parameter family of solutions  (see  Eq. (C.7) of \cite{chipum2})
\be \label{Glambda}
\G_\lambda(x,x') =  - \frac{1}{4\pi} \left(   \frac{2\chi^2-1}{\sqrt{\chi^2-1}} + \lambda \chi\right) 
\ee
where
\be
\chi = - V(x) \cdot V(x')= \cosh d(x,x')
\ee
and  $\lambda$ is an undetermined constant. 
For a given choice of this constant, the resulting potential is then
\ba \label{sigmalambdasum}
\sigma^{(\lambda)}(x) &= & -G \sum_{\substack{i \\ m_i \neq 0}}  m_i \left(   \frac{2\chi^2_i-1}{\sqrt{\chi^2_i-1}} + \lambda \chi_i \right) \\
&= & -G \sum_{\substack{i \\ m_i \neq 0}}     \frac{2(V \cdot p_i)^2-m_i^2}{\sqrt{(V \cdot p_i)^2-m_i^2}} + G \lambda V \cdot P_\massive \label{sigmalambdaitop}\\
\ea
where
\be
V^\mu \equiv  V^\mu(x), \quad V^\mu_i \equiv V^\mu(x_i), \quad p^\mu_i \equiv m_i V^\mu_i ,
\ee
and
\be
P^\mu_\massive = \sum_{\substack{i \\ m_i \neq 0}} p^\mu_i
\ee
is the total  momentum of the massive particles. Comparing with \eqref{delLsigma}, we see that different choices of $\lambda$ are related to each other by log translations that are proportional to $P^\mu_\massive$.

A natural way to eliminate the log translation freedom is to require that $\sigma$ vanishes asymptotically \cite{cgw}
\be\label{asymsigmacgw}
\lim_{\rho \to \infty}\sigma = 0 \iff  c^\mu_\scri=0  \quad \text{(radiative log frame)},
\ee
which  amounts to choosing $\lambda=-2$ in \eqref{sigmalambdasum}.  This condition singles out a specific log translation frame that we refer to as radiative frame.\footnote{As discussed in  subsection \ref{globallogsec}  there are in fact two distinct radiative frames, depending on whether condition \eqref{asymsigmacgw} is imposed at the future or at the past.}

\subsubsection*{``Bulk''  approach in harmonic gauge} \label{harmggesec}

In harmonic  gauge, the spacetime metric is written as
\be \label{gharm}
g^\harm_{\mu \nu} = \eta_{\mu \nu}(1-e) +2 e_{\mu \nu},
\ee
where $\eta_{\mu \nu}$ is a reference  Minkowski metric,  with associated Cartesian coordinates $X^\mu$.   
$e_{\mu \nu}$  is the so-called trace-reversed metric perturbation and $e = \eta^{\mu \nu} e_{\mu \nu}$ its trace. The gauge condition is  
\be \label{ggecond}
\partial_\nu  e^{\mu \nu} =0,
\ee
and allows Einstein equations to be written as
\be \label{boxEmunu}
\square e_{\mu \nu}= -8 \pi G \T_{\mu \nu} , \quad \T_{\mu\nu} = T_{\mu \nu} + T^\grav_{\mu \nu},
\ee
where $\square \equiv \partial^\mu \partial_\mu$ is the d'Alembertian wrt the flat metric,  $T_{\mu \nu}$ is the usual matter stress-energy tensor  and $T^\grav_{\mu \nu}$  is a ``gravitational stress-energy tensor'' that is   constructed out of $e_{\mu \nu}$ and its derivatives.  We emphasize that \eqref{boxEmunu} is simply a  rewriting of the full non-linear Einstein equations \cite{senreview}. 

The formal solution to \eqref{boxEmunu} with retarded boundary conditions is
\be \label{4dgrn}
e_{\mu \nu}(X) = 4  G  \int d^4 X'  \theta(X^0-{X'}^0) \delta((X-X')^2) \T_{\mu\nu}(X'),
\ee
and can be  evaluated perturbatively by regarding $e_{\mu \nu}$ as a series expansion in $G$. However, for the purposes of obtaining the leading late-time component 
 \be \label{Eone}
e_{\mu \nu} \stackrel{\t \to \infty}{=} \frac{1}{\t} \ov{1}{e}_{\mu \nu}(x) + \cdots, 
\ee
it suffices to evaluate \eqref{4dgrn} at the zeroth perturbative order,   for which the only contribution to $\T_{\mu\nu}$ is due to the freely propagating particles, see e.g. \cite{laddhasenlast}.  Setting
 \be
X'^\mu=\t' V^\mu(x')+ \cdots, \quad  X^\mu = \t V^\mu(x) + \cdots, \quad \t \to \infty
  \ee
 in \eqref{4dgrn} leads to (see appendix A of \cite{logwalok} for an analogous computation)
 \be \label{Eoneint}
\ov{1}{e}_{\mu \nu}(x) = 2 G \int \frac{d^3 x'}{\sqrt{\chi^2-1}} V_\mu(x') V_\nu(x') \, \rho(x')
\ee
where
\be \label{fullrho}
\rho(x')=\sum_i m_i \delta(x',x_i)
\ee
is the energy density at timelike infinity \emph{including} massless contributions, as we discuss below.

We now extract from \eqref{Eoneint} the gravitational potential $\sigma$. The $O(1/\t)$ metric perturbation \eqref{defgamma} is  given by $\gamma_{\mu \nu}=  2 \ov{1}{e}_{\mu \nu}- \ov{1}{e} \eta_{\mu \nu}$, and  from the definition of $\sigma $ \eqref{defsigma}  one gets
\ba
 \sigma^\harm &=& - \ov{1}{e}_{\mu \nu} V^\mu V^\nu  -\frac{1}{2}\ov{1}{e}  \\
 &=&  - G \int d^3 x' \frac{2\chi^2 -1}{\sqrt{\chi^2-1}} \, \rho(x').
\ea

Upon using \eqref{fullrho},  this yields a potential that looks like $\sigma^{(\lambda=0)}$ in \eqref{sigmalambdasum}, but where the sum includes both massive and  massless particles. To account for the latter, we express the potential in terms of the particle momenta, as in \eqref{sigmalambdaitop}, leading to,
\be \label{sigmaharmitop}
 \sigma^\harm  =   -G \sum_{i}     \frac{2(V \cdot p_i)^2-m_i^2}{\sqrt{(V \cdot p_i)^2-m_i^2}} 
 = \sigma^{(\lambda=0)} + 2 G V \cdot P_\massless ,
 \ee
where the two terms in the last equality  correspond to the contributions from massive and massless particles
to the sum, and where we used that  $|V \cdot p_i|=-V \cdot p_i$.

On the other hand, 
according to the discussion from the previous section,  the potential in the radiative log frame \eqref{asymsigmacgw}
 can be written as
\be \label{sigmabs}
\sigma^\rad =  \sigma^{(\lambda=-2)}=\sigma^{(\lambda=0)} - 2 G  V \cdot P_\massive .
\ee
Solving for $\sigma^{(\lambda=0)}$ in \eqref{sigmabs} and substituting in \eqref{sigmaharmitop} leads to
\be
 \sigma^\harm = \sigma^\rad +2 G  V \cdot P_\total,
\ee
where $P^\mu_\total=  P^\mu_\massive +  P^\mu_\massless$.

Thus, the harmonic and radiative frame potentials are related by a log translation that is proportional to the total spacetime momentum. 
In particular, the null limit of the log deviation vector is now given by
\be\label{asymsigmaharm}
 c^\mu_\scri= 2 G   P^\mu_\total \quad \text{(harmonic log frame)}. 
\ee
This reproduces, in our conventions, Eq. (3.29) of \cite{sahoosen}. The expression is also consistent with the well-known asymptotic diffeomorphism  that interpolates between harmonic and radiative coordinates, see e.g. Eq. (95) of \cite{blanchetreview}. We refer to \eqref{asymsigmaharm} as the harmonic log frame.\footnote{We are here presenting the  harmonic log frame as a condition on the log deviation vector of outgoing null rays. One can alternatively write it as a condition on incoming null rays, as reviewed in the next  subsection.  There is  a non-trivial relationship between the future and past null deviation vectors which we discuss in  subsection \ref{globallogsec}.}

\subsection{Future vs. past}\label{pasttimesec}
The discussion of the previous subsections has a direct analogue at past timelike infinity.  Below we summarize the equations and conventions that we shall use to describe both infinities. 
When needed, we will distinguish future and past quantities by a $+/-$ label.

\begin{itemize}
\item Asymptotic geodesics:

\be \label{Xmulambdaplusminus}
X^\mu(s) \stackrel{s \to \pm \infty}{=} s V^{\mu}_\pm + \log | s | c^\mu_\pm + \cdots, \quad   \text{$V^\mu_\pm$ future-pointing}. 
\ee

\item Spacetime hyperbolic coordinates: $(\t>0,x^a)$ and $(\t<0,x^a)$ with
\be 
X^\mu \stackrel{\tau \to \pm \infty}{=} \tau V^\mu(x) + \cdots, \quad V^\mu V_\mu = -1, \quad \text{$V^\mu$ future-pointing}.
\ee

\item Gravitational potential and log deviation vector:
\be
\sigma_\pm =  -\tfrac{1}{2} \lim_{\t \to \pm \infty}   V^\mu V^\nu |\t| (g_{\mu \nu} - \eta_{\mu \nu}) 
\ee
\be \label{cpmitosigma}
c^\mu_\pm = \pm( D^a V^\mu \partial_a \sigma_\pm - V^\mu \sigma_\pm).
\ee
\be
\sigma_\pm =  \pm c^\mu_\pm V_\mu.
\ee
\be \label{cmuscriti}
c^\mu_{\scri^\pm} = \lim_{\rho \to \infty} c^\mu_\pm
\ee

\item Energy density and leading Einstein equation:
\be
\rho_\pm = \lim_{\t \to \pm \infty} |\t^3| V^\mu V^\nu T_{\mu \nu}
\ee
\be  \label{EEsigmapm}
(D^2-3) \sigma_\pm = 4 \pi G \rho_\pm
\ee

\item Gravitational potential in terms of outgoing/incoming momenta:
 
We follow standard amplitude conventions in which the particles' momenta at past timelike infinity are taken with an overall minus sign (so that  they point to the past): 
\be\label{defpinout}
p^\mu_{i} = \pm m_i V^{\mu}_{\pm i}.
\ee

With these conventions, the gravitational potential at future/past timelike infinity in the various log frames discussed earlier is 
\ba
\sigmazero_\pm &=&  -G \sum_{\substack{i \in \pm \\ m_i \neq 0}}     \frac{2(V \cdot p_i)^2-m_i^2}{\sqrt{(V \cdot p_i)^2-m_i^2}} \label{sigmazero} \\
\sigma^{\rad \pm}_\pm &=& \sigmazero_\pm \mp 2 G V \cdot P_{\massive \pm}\\
\sigma^\harm_\pm &=&  \sigma^\rad_\pm \pm 2 G  V \cdot P_{\total \pm}
\ea
where $\sigmazero_\pm \equiv \sigma^{(\lambda=0)}$, $i\in \pm$  indicates outgoing/incoming particles and  ``$\rad \pm$'' refers to the  radiative log frame at future/past infinity. $P^\mu_{\total \pm}= P^\mu_{\massive \pm} +  P^\mu_{\massless \pm}$ is the total outgoing/incoming momentum, with the conventions of \eqref{defpinout} so that momentum conservation reads 
\be \label{consP}
P^\mu_{\total +}+P^\mu_{\total -}=0.
\ee

Since $P^\mu_{\total +}$ is just the total momentum of the system,  we shall occasionally omit the plus subscript and write it as
\be
P^\mu_{\total } \equiv  P^\mu_{\total +} .
\ee

\item The  log deviation vector can be obtained by applying \eqref{cpmitosigma} to the corresponding gravitational potential. In particular, the deviation vector associated to the $\lambda=0$ potential \eqref{sigmazero} is
\be \label{czeropm}
\czero^\mu_\pm = \mp G  \sum_{\substack{i \in \pm \\ m_i \neq 0}}     \frac{(2 (V \cdot p_i)^3-3 m_i^2 V \cdot p_i )p_i^\mu -m^4_i V^\mu}{\left((V \cdot p_i)^2-m_i^2\right)^{3/2}} .
\ee
The  deviation vector associated to the other potentials can be obtained from \eqref{czeropm} by appropriate log translations. In the   harmonic log frame the expression  has the form \eqref{czeropm} but with the sum running over all (massive and massless) outgoing/incoming particles. In particular, one finds\footnote{The sign difference in \eqref{czeropm} disappears in \eqref{harmcpm} because  $|V \cdot p_i|=\mp V \cdot p_i$ for outgoing/incoming $p_i$, according to the conventions   \eqref{Xmulambdaplusminus} and \eqref{defpinout}.}
\be \label{harmcpm}
    c^\mu_{\scri^\pm}=2 G   P^\mu_{\total \pm}  \quad \text{(harmonic log frame)}.
\ee

\end{itemize}

\subsection{Global log translations} \label{globallogsec}

A priori, there are two independent log translations at the future/past
\be
\xi^\mu_{L^\pm} \stackrel{\t \to \pm \infty}{=} \log |\t|  L^\mu_\pm + O(\t^0),
\ee
which act according to
\be
\delta_{L_\pm} c^\mu_\pm = - L^\mu_\pm, \quad\delta_{L_\pm} \sigma_{\pm} = \mp L^\mu_\pm V_\mu .
\ee

However, as in the case of ordinary translations (or more generally,  BMS supertranslations \cite{stromgravscatt}), these two transformations should not be regarded as independent, but instead  satisfy
\be\label{LpluseqLminus}
L^\mu_+=L^\mu_- \equiv L^\mu.
\ee
This condition arises naturally when working in harmonic coordinates, where a single patch covers all spacetime (at least outside the scattering bodies). Alternatively, as discussed in \cite{gianni2}, it follows from the matching properties of the gravitational field across timelike, null and spatial infinity.

The identification \eqref{LpluseqLminus} leads to a single log translation group acting simultaneously on both future and past log deviation vectors,
\be \label{delLallcmu}
\delta_{L} c^\mu_\pm = - L^\mu.
\ee
This allows for a global notion of log frame, i.e., one that applies to all infinities.  Among the possible choices, three global log frames are of special relevance: the future and past radiative  frames, and the harmonic frame.

The first two  are characterized by the vanishing of the corresponding log deviation vector of null geodesics,
\be \label{radpmlogframe}
 c^\mu_{\scri^\pm}=0 \quad  \text{(rad$^\pm$ log frame)}.
\ee

The harmonic log frame corresponds to  condition \eqref{harmcpm} on either outgoing or incoming null rays. In view of  \eqref{LpluseqLminus}, however, the two conditions cannot be regarded as independent. Taking the sum and difference of  \eqref{harmcpm} and using momentum conservation one finds,
\ba \label{harmggesymprel}
 c^\mu_{\scri^+}+c^\mu_{\scri^-} &= &0 \quad \text{(harmonic log frame)} \\
c^\mu_{\scri^+}-c^\mu_{\scri^-}&=& 4 G P^\mu_{\total} \label{cplusminuscminus} .
\ea
Only the first combination is not invariant under  \eqref{delLallcmu}, and thus serves to fix global  log translations.   By contrast, Eq. \eqref{cplusminuscminus}  is an identity that  holds in any global log frame, even though it was obtained  from a harmonic gauge computation.\footnote{In particular, we learn that in the future  radiative frame    $c^\mu_{\scri^-}=- 4 G P^\mu_{\total} $, while in the past radiative frame  $c^\mu_{\scri^+}= 4 G P^\mu_{\total} $.}  \\

\noindent {\bf Comment} \\
\noindent  It is interesting to note that there is a spatial infinity version of these equations, in which  $c^\mu_{\scri^\pm}$ is obtained from the null limit of asymptotic spacelike geodesics. In that context, \eqref{harmggesymprel} coincides with the  log translation fixing proposed  in~\cite{aalog}, while \eqref{cplusminuscminus}  reproduces a long-known ``discontinuity" of the gravitational field  at spatial infinity~\cite{hansen,tn,cgw}. See \cite{gianni2} for further details. 

\newpage

\section{Soft theorems and their log translation invariance} \label{logsoftsec}

The classical soft theorems found in \cite{laddhasen1,laddhasen,sahoosen,proofdeq4,sahoosubsub}  constrain the soft limit of the gravitational waveform through certain universal terms.  Specifically, the Fourier transform of the  metric perturbation at null infinity satisfies
\begin{equation} \label{hmunuomega}
\tilde{h}_{\mu\nu}(r,\omega, \hat{n}) \stackrel{\w \to 0}{=} \frac{1}{r} \left(  \omega^{-1} h^{(0)}_{\mu\nu}(\hat{n}) + \log \omega \, h^{(\log)}_{\mu\nu}(\hat{n}) +  \w \log^2 \w \, h^{(\log^2)}_{\mu\nu}(\nh)+ \cdots \right)
\end{equation}
where  we only kept the $O(1/r)$ radiative field, $\w$ is the frequency and $\hat{n}$ the direction on the celestial sphere from which the waveform is observed.    The leading   term\footnote{Here and in what follows  the index $i$ runs over  all incoming and outgoing particles, with  sign conventions as in \eqref{defpinout}, and  $n^\mu$ is the future-pointing null direction associated to $\nh$  \eqref{defnmu}.  The sums  include both massive and massless ``particles'', the latter capturing the contribution from radiation (gravitational or other) \cite{senreview}.}
\be \label{hminuone}
h^{(0)}_{\mu\nu}(\hat{n})= -  4 G i \sum_i \frac{p^i_\mu p^i_\nu}{p_i \cdot n},
\ee
 is described  by Weinberg's soft  theorem \cite{weinberg} and captures what is known as the gravitational memory effect  \cite{zhibo}.     The results from  \cite{laddhasen1,laddhasen,sahoosen,proofdeq4,sahoosubsub}  concern the  subleading logarithmic terms. In this section we focus on the  first two of what are believed to be infinitely many universal ``leading logs''  $\w^{k-1} \log^{k} \w$ with  $k=1,2,\ldots$ \cite{proofdeq4,sahoosubsub,rewritten,spectra,senreview}.\footnote{See \cite{Ghosh:2021bam,Akhtar:2024lkk} for results on non-leading soft logs.}

The proofs of these soft theorems are based on rather involved perturbative computations. There is, however, a shortcut that reproduces these results more directly~\cite{laddhasen1,sahoosen}. Although it does not amount to a full proof, it provides insight into the origin of the various contributions. More importantly for our purposes, it reveals where the harmonic log frame is implicitly assumed and thus how it can be relaxed.

Following  section 5 of \cite{proofdeq4}, the starting point for the shortcut  derivation is the tree-level (or $d>4$) $O(G)$ soft expansion \cite{stromingercachazo,Laddha:2018rle,laddhasenlast,laddhasensubsub}
\begin{equation}  \label{treesoftexp1st2}
\tilde{h}^\tree_{\mu\nu}(r,\omega, \hat{n}) \stackrel{\w \to 0}{=}   \frac{1}{r}  \left(  \omega^{-1} h^{(0)}_{\mu\nu}(\hat{n}) + h^{(1)}_{\mu\nu}(\hat{n})  + \w \, h^{(2)}_{\mu\nu}(\hat{n}) +  \cdots \right) 
\end{equation}
where\footnote{There could be  non-universal terms at order $O(\w)$ \cite{laddhasensubsub} which however play no role for the log soft theorems.} 
\be \label{treelevelsubfactors}
h^{(1)}_{\mu\nu} =  4G \sum_i \frac{p^i_{(\mu}  J^{i}_{\nu)\rho} n^{\rho}}{p_i \cdot n},  \quad h^{(2)}_{\mu\nu} =  2 i G  \sum_i \frac{ J^{i}_{\mu \rho}J^{i}_{\nu \sigma}  n^{\rho} n^{\sigma}}{p_i \cdot n},
\ee
with $J^{i}_{\mu \nu}$  the particle angular momenta.

The expansion \eqref{treesoftexp1st2}  ignores the effect of the logarithmic deviation in the asymptotic trajectories. This can be incorporated by taking into account: 1) The logarithmic deviation of the soft null rays and 2) The logarithmic deviation of the hard particles. The first effect, referred to as gravitational drag, is incorporated via the phase factor \cite{laddhasen1,sahoosen}
\be \label{dragphase}
\tilde{h}^\tree_{\mu\nu} \rightarrow  e^{- i \w \log \w  \, n \cdot c_{\scri^+}} \tilde{h}^\tree_{\mu\nu},
\ee
where, in harmonic gauge,   $c^\mu_{\scri^+} = 2 G P^\mu_{\total}$, see  Eq. \eqref{asymsigmaharm}.

The second effect is due to the divergent nature of the particle angular momenta. For asymptotic trajectories of the form \eqref{Xmulambdaplusminus} one has
\begin{equation} \label{Jmunudiv}
J_{\mu\nu}^i   \stackrel{s \to \pm \infty}{=}   \log | s | J^{\div \, i}_{\mu \nu} + \cdots, 
\end{equation}
where
\be \label{defJdiv}
J^{\div \, i}_{\mu \nu} :=  c^i_\mu p^i_\nu - c^i_\nu p^i_\mu .
\ee
The corresponding contribution to the log soft theorem is then obtained by substituting \eqref{Jmunudiv} in \eqref{treesoftexp1st2} and making the identification \cite{laddhasen1}
\be \label{laddhasenidentification}
\log |s| \sim  \log (\omega^{-1}).
\ee

Including both effects, one arrives at \cite{proofdeq4}
\begin{equation} \label{finalstepshortcut}
\tilde{h}_{\mu\nu}= \frac{1}{r} e^{- i \w \log \w  \, n \cdot c_{\scri^+}} \left[  \omega^{-1} h^{(0)}_{\mu\nu} - \log \w\, h^{(1) \div}_{\mu\nu}  + \w \log^2 \w \, h^{(2) \div}_{\mu\nu} +  \cdots \right],
\end{equation}
where the ``div'' label indicates that  $J_{\mu\nu}^i $ is replaced by $J^{\div \, i}_{\mu \nu}$ in the expressions \eqref{treelevelsubfactors}.

Finally, by expanding the exponential in \eqref{finalstepshortcut}, one recovers \eqref{hmunuomega} with
\ba
 h^{(\log)}_{\mu\nu} &=& - i  n \cdot c_{\scri^+} \, h^{(0)}_{\mu\nu} -  h^{(1) \div}_{\mu\nu}  \label{hlog},\\
h^{(\log^2)}_{\mu\nu}&= &-\tfrac{1}{2} (n \cdot c_{\scri^+})^2 \, h^{(0)}_{\mu\nu}  + i  n \cdot c_{\scri^+} h^{(1) \div}_{\mu\nu} + h^{(2) \div}_{\mu\nu}
\ea

Remarkably, these expressions correctly reproduce the rigorously  derived log soft factors \cite{proofdeq4,sahoosubsub}. We recall that, in both the heuristic and rigorous derivations, the deviation vectors  are  written in the harmonic log frame. We now show that the above expressions are in fact valid in any log frame.

Under a log translation, both the ``soft'' null-ray deviation vector in \eqref{dragphase}  and the ``hard'' particles' deviation vector in \eqref{defJdiv} transform according to  \eqref{delLallcmu}
\be \label{delLcmusoftthm}
 \delta_L c^\mu_{\scri^+}= \delta_L c^\mu_i =-L^\mu .
\ee
Applying this to the factors $h^{(1) \div}_{\mu\nu}$ and $h^{(2) \div}_{\mu\nu}$ one finds 
\be
\delta_L h^{(1) \div}_{\mu\nu} =  i L \cdot n \,  h^{(0)}_{\mu\nu}, \quad     \delta_L h^{(2) \div}_{\mu\nu} =  i L \cdot n \,  h^{(1) \div}_{\mu\nu}\label{delLhs}     
\ee
where in the first relation  we used momentum conservation $\sum_i p^i_\mu=0$ and in the second one  $\sum_i J^{\div \, i}_{\mu \nu} =0$.\footnote{In fact, the total incoming and outgoing divergent angular momentum vanishes separately.}  
Using \eqref{delLcmusoftthm} and \eqref{delLhs} one can  verify the invariance of the log soft factors
\be
\delta_L h^{(\log)}_{\mu\nu} =0, \quad  \delta_L  h^{(\log^2)}_{\mu\nu} =0  .
\ee

 The result just   proven can be used to explain the   cancellations among  outgoing massless particles observed in \cite{sahoosen,rewritten}. In  the harmonic gauge setting used in these references, the dependence on outgoing massless particles appears through:  1) their total  momentum in the drag term \eqref{dragphase} and 2) their individual  momenta in the factors $h^{(k) \div}_{\mu\nu}$. The cancellation of these two contributions can be traced back to  the fact that  $c_i^\mu \equiv c^\mu_{\scri^+}$ for outgoing massless particles. 

The independence   from outgoing massless particles is however manifest  if one writes the log soft factors in the future radiative log frame, where $c^\mu_{\scri^+}=0$. In that case, each contribution vanishes separately.\\

\noindent {\bf Comments}
\begin{itemize}

\item As a consistency check and illustration,  in  appendix~\ref{recoverysahoosenapp} we  evaluate the first log term \eqref{hlog} both in harmonic and radiative log frames,  recovering the expression as presented in \cite{sahoosen,senreview}.

\item In time domain, the soft expansion translates into early and late time tails in the gravitational waveform that can be extracted by giving the soft frequency a small positive/negative imaginary part \cite{senreview}.  Without such prescription, the expansion \eqref{hmunuomega} only captures \emph{differences} between these early/late time contributions, see appendix B of \cite{spectra} for further details. The soft terms considered here and in \cite{gianni2}  are  associated to these differences.  In particular, the log translation invariance does not apply to individual early/late time components.

\item Invariance under log translations leads to non-trivial identities only when the soft factors are expressed in terms of the deviation vectors. Once the deviation vectors are written in terms of particle momenta (in any log frame),  the invariance is trivially satisfied, since particle momenta are unaffected by log translations.

\item   The validity of \eqref{finalstepshortcut} for the second log relies on the fact \cite{proofdeq4,sahoosubsub} that the subleading deviation in the asymptotic trajectory \eqref{Xmulambda} is of order $\ln s / s$, and therefore does not contribute to $\w \log^2 \w$ in the replacement $s \to \w^{-1}$. This fall-off can be shown by expanding \eqref{4dgrn} to subleading order, which yields an $O(\ln \t / \t^2)$ term.

\end{itemize}

\section{On higher order soft theorems}   \label{higherordersec}
Following the success of the heuristic derivation of the first two log soft theorems, the general structure of higher-order leading log terms was described in \cite{sahoosubsub,spectra}. In this section we revisit these considerations from the perspective  of the preceding discussion.

The starting point is again  the $O(G)$ waveform. One can argue on general grounds that it  admits a power expansion in the frequency that extends \eqref{treesoftexp} to all orders
\begin{equation}  \label{treesoftexp}
\tilde{h}^\tree_{\mu\nu}(r,\omega, \hat{n}) =   \frac{1}{r\w}  \sum_{k=0}^\infty  \omega^{k} h^{(k)}_{\mu\nu}(\hat{n}) .
\end{equation}
In \cite{hamada,zhang} it was shown that the  $k \geq 2$  coefficients take the form\footnote{The analysis of  \cite{zhang,hamada} is in the context of momentum-space scattering amplitudes. We adapt their expressions by making the replacement $\partial/\partial p^i_\mu \rightarrow -i b^\mu_i$.}
\be \label{hktree}
h^{(k)}_{\mu\nu}(\hat{n}) =  -4 G i \frac{(-i)^{k}}{k!} \sum_i \frac{ J^{i}_{\mu \rho}J^{i}_{\nu \sigma}  n^{\rho} n^{\sigma}}{p_i \cdot n} (b_i \cdot n)^{k-2} + \Rt^{(k)}_{\mu \nu}, \quad k \geq 2,
\ee
for  $O(G^0)$ asymptotic trajectories $X^\mu_i(s)=\pm s p^\mu_i/m_i +b_i + \cdots$.  These are referred to as partial soft theorems, since there is an undetermined piece $\Rt^{(k)}_{\mu \nu}$. The  ``universal'' term in \eqref{hktree} is distinguished from this remainder through its  dependence on  $n^\mu$, the latter being  of the form $\Rt^{(k)}_{\mu \nu} = \Rt^{(k)}_{\mu \nu \, \alpha_1 \ldots \alpha_{k-1}} n^{\alpha_1} \ldots n^{\alpha_{k-1}}$. The remainders are generically non-trivial, except for the $k=2$ case which vanishes according to  \eqref{treelevelsubfactors}.

Doing the replacement  $b^\mu_i \rightarrow - \ln \w c_i^\mu$,  one arrives at the expansion \cite{sahoosubsub,spectra}
 \be \label{allorderslogs}
\tilde{h}_{\mu\nu}(r,\omega, \hat{n}) \approx -\frac{4 G i}{r \w} e^{- i \w \log \w  \, n \cdot c_{\scri^+}} \sum_{k=0} \frac{1}{k!} (- i \w  \log \w )^k  \, a^{(k)}_{\mu \nu} ,
 \ee
 where the wiggly equal sign means that we are ignoring terms of the form $\w^k \log^n \w$ with $n<k$ and where we adopt the normalization of \cite{spectra} for the coefficients.\footnote{Except for the $k=0$ one, which here differs by an overall sign from \cite{spectra}. To relate their expressions with ours, note that the log deviation vector of the $i$-th particle in harmonic gauge can be written as $c^\mu_i = - 2 G \sum_j \t_{i j} p^\mu_j+ (\cdots) p^\mu_i$ where the sum is over either incoming or outgoing particles and $\t_{ij}$ is the ``relative logarithmic drift'' as  defined in \cite{spectra}.} These are given by
 \be \label{a01}
a^{(0)}_{\mu\nu}=  \sum_i \frac{p^i_\mu p^i_\nu}{p_i \cdot n}, \quad a^{(1)}_{\mu\nu} = \sum_i \frac{p^i_{(\mu}  J^{\div \, i}_{\nu)\rho} n^{\rho}}{p_i \cdot n},  
\ee
\be \label{ansatzak}
 a^{(k)}_{\mu \nu} = (-1)^k  \sum_i \frac{ J^{\div \, i}_{\mu \rho}J^{\div \, i}_{\nu \sigma}  n^{\rho} n^{\sigma}}{p_i \cdot n} (c_i \cdot n)^{k-2} + r^{(k)}_{\mu \nu}, \quad k \geq 2,
 \ee
with $r^{(k)}_{\mu \nu}= r^{(k)}_{\mu \nu \, \alpha_1 \ldots \alpha_{k-1}} n^{\alpha_1} \ldots n^{\alpha_{k-1}}$ undetermined for $k \geq 3$.

In \cite{spectra}, a proposal is given for the remainders whose  general applicability is yet to be established \cite{senreview}. Unfortunately, the proposal of \cite{spectra} does not appear to admit a rewriting in terms of deviation vectors and hence we cannot establish its consistency with log translation invariance.

We can, however, reverse the logic of the previous section and ask: if the remainders admit arbitrary log-frame expressions, what restrictions are imposed by log translation invariance? To answer this question, we first note that, by applying an infinitesimal log translation to \eqref{allorderslogs} and requiring it to vanish leads to
\be \label{recursionak}
\delta_L  a^{(k)}_{\mu \nu} = k \, (n \cdot L) \, a^{(k-1)}_{\mu \nu},
\ee
generalizing the relations \eqref{delLhs} to arbitrary order.  Using the general form \eqref{ansatzak} in \eqref{recursionak} leads to, 
\begin{multline} \label{recursionrk}
\delta_L  r^{(k)}_{\mu \nu} =  k \, (n \cdot L) \, r^{(k-1)}_{\mu \nu}  \\
-  (-1)^{k} \sum_{i} \left( L_{[\mu} c^i_{\rho]} p^i_{[\nu} c^i_{\sigma]} + (\mu \leftrightarrow \nu) \right)n^\rho  n^\sigma (c_i \cdot n)^{k-3} , \quad k \geq 3, 
\end{multline}
where $v_{[\mu} w_{\nu]} \equiv v_\mu w_{\nu}- v_\nu w_\mu$. We emphasize the non-trivial cancellation in the $1/(p_i \cdot n)$ dependence  on both sides of \eqref{recursionak}, so that the recursion relation for the remainder is 
polynomial in $n^\mu$.

We leave for future investigations the study of solutions of \eqref{recursionrk}. The initial seed is given by $ r^{(2)}_{\mu \nu} =0$. Although not immediately obvious, the integrability of \eqref{recursionrk} (i.e. the condition that the RHS of \eqref{recursionrk} is consistent with $[\delta_L,\delta_{L'}]=0$) follows from that of \eqref{recursionak}, where it is trivially satisfied. Note also that \eqref{recursionrk} is consistent with ordinary gauge invariance, $r^{(k)}_{\mu \nu} n^\nu=0$.

\section{Discussion} \label{finalsec}

In this paper, we uncovered the role of log translations in classical gravitational soft theorems. Unlike the by now standard link between asymptotic symmetries and soft theorems, log translations are pure gauge and lead to trivial conservation laws.   However, by writing the soft factor in terms of log translation–dependent quantities  one can extract non-trivial information on the form of these coefficients. In particular, this explains why  the  log soft factors cannot depend on outgoing massless particles.

We explicitly checked the log translation invariance of the first two known log soft coefficients. Although there exist specific proposals for higher order terms, these do not admit any obvious rewriting in terms of log translation–dependent quantities and hence we could not verify their consistency with log translations.  Nevertheless, if general log frame expressions exist for such terms, they would  obey a recurrence relation that constrains their structure.

From a broader perspective, our results represent a new example, among  the many that followed \cite{stromgravscatt}, of the rich interplay between perturbative results and the geometric description of the gravitational field at infinity.
In an upcoming  paper \cite{gianni2} we deepen this interplay by providing a proof of the first log soft theorem purely from an asymptotic perspective.

There are many avenues for future research. It would be important to understand how the considerations presented here connect to the realization of soft theorems as Ward identities. These are typically discussed in a context where log translations are fixed (e.g. by working in radiative or harmonic coordinates). A first step in this direction would be to achieve a description of superrotations (and, more generally, asymptotic higher-spin symmetries \cite{Guevara:2021abz,stromwinf,Freidel:2021ytz,geillerwinfty,Cresto:2024fhd,Cresto:2024mne}) that is valid in arbitrary log frames. Such a framework should be useful for an eventual extension of \cite{chipum2} to the higher-order log soft theorems.

Our focus was on the classical theory, but much of the motivation comes from quantum gravity. It would be interesting to explore the consequences of log translation invariance for the quantum soft theorems \cite{Bern:2014oka,He:2014bga,sahoosen,Krishna:2023fxg}, and, more ambitiously, for asymptotically flat holography \cite{Susskind:1998vk,solo,Marolf:2006bk,Barnich:2009se,raju,Raclariu:2021zjz,Bagchi:2023cen,Sen:2025oeq}.

 \acknowledgments
  We would like to thank Guzmán Hernández-Chifflet for collaboration during the initial stages of this project and to Alok Laddha for emphasizing to us the  puzzling cancellations.  We thank Glenn Barnich, Federico Capone, Geoffrey Compère, Alok Laddha, Carlo Heissenberg, Guzmán Hernández-Chifflet and Sébastien Robert for fruitful discussions.   GB would like to thank  the ULB theory group for hospitality during the completion of this work.  MC  thanks the organizers and participants of the GGI workshop \emph{From Asymptotic Symmetries to Flat Holography: Theoretical Aspects and Observable Consequences} for a stimulating environment where a preliminary version of this work was presented.   We acknowledge  support from Pedeciba and from ANII grants POS-NAC-2023-1-177577 and FCE-1-2023-1-175902.

\appendix

\section{More on timelike infinity}

In this appendix, we expand the discussion of timelike infinity given in the body of the paper. We start by introducing notation and summarizing useful identities. We then present a systematic description of the first order gravitational field at timelike infinity, and specialize it to the harmonic gauge case. In this setting, we give a general argument showing that log translations are frozen, and  recover the  BMS diffeomorphisms at timelike infinity as described in  \cite{clmassive}. Finally, we show that the metric \eqref{4dgrn} has a vanishing asymptotic magnetic Weyl tensor, consistent with standard assumptions in the asymptotic literature.

\subsection{Hyperboloid decomposition of tensors} \label{hyperapp}
Given a Minkowski 4-vector \(F^\mu\), we write its decomposition with respect to the hyperbolic splitting as
\be
F^\mu = f V^\mu +f^a \partial_a V^\mu
\ee
where
\be \label{fitoF}
f = - V_\mu F^\mu,  \quad  f^a = D^a V_\mu  F^\mu.
\ee

We note the identities:
\begin{align}
F_\mu G^\mu & = - f g + f^a g_a  ,\\
D^a V_\mu D_a V_\nu &= \eta_{\mu \nu} + V_\mu V_\nu .
\end{align}

More generally, one  can decompose any Minkowski tensor into longitudinal and transverse components by contracting its Lorentz indices with $\eta_{\mu \nu} = - V_\mu V_\nu + D^a V_\mu D_a V_\nu$. For instance, a symmetric tensor \(T_{\mu \nu}\) can be written as
\be \label{hypdectensor}
T_{\mu \nu} = t_\parallel V_\mu V_\nu - 2 t_a D^a V_{(\mu} V_{\nu)} + t_{ab} D^a V_\mu D^b V_\nu
\ee
with
\bal
t_\parallel &= V^\mu V^\nu T_{\mu \nu} ,\\
t_a &= D_a V^\mu V^{\nu} T_{\mu \nu}, \\
t_{ab} &= D_a V^\mu D_b V^\nu T_{\mu \nu}.
\eal

The subscript in the 3-scalar is to distinguish it from the trace of the 3-tensor,
\be
t_\perp := h^{ab} t_{ab}
\ee
so that
\be
T \equiv \eta^{\mu \nu} T_{\mu \nu} = - t_\parallel + t_\perp.
\ee

\subsubsection*{Useful identities} \label{usefulids}
\be
D_a D_b V^\mu = h_{ab} V^\mu,
\ee
\be
[D_a,D_b] f_c = f_a h_{bc} -f_b  h_{ac}.
\ee

\subsection{Asymptotic diffeos and asymptotic Weyl tensor} \label{timeinfapp}
In this section we present several details of the discussion given in section \ref{timeinfsec}. Our starting point is the metric \eqref{defgamma}
 \begin{equation} \label{defgammaapp} 
g_{\mu \nu} = \eta_{\mu \nu} + \frac{1}{\t} \gamma_{\mu \nu} + \cdots.
\end{equation}
The asymptotic vector fields preserving this form are given by
\be \label{xiapp}
\xi^\mu = \t R^{\mu \nu} V_\nu + \log \t L^\mu + F^\mu + \cdots,
\ee
where $R^{\mu \nu}$ is a constant antisymmetric matrix representing Lorentz transformations and $L^\mu$ a constant vector representing log translations.  $F^\mu$, on the other hand, depends on the hyperboloid point. Within it, one finds regular translations, supertranslations, and pure gauge transformations, as we discuss below. From now on we set $R^{\mu \nu}=0$;  its only effect at this order being a Lorentz rotation on $\gamma_{\mu \nu}$. 

To evaluate the action of the remaining components in \eqref{xiapp}, we write
\be \label{Liegapp}
\Lie_{\xi} g_{\mu \nu} = 2 \partial_{(\mu} \xi_{\nu)} + \cdots
\ee
where the dots indicate subleading terms that do not affect $\gamma_{\mu \nu}$. The  Cartesian derivatives in terms of hyperbolic coordinates are given by
\be \label{partialmuitohyp}
\partial_\mu  = - V_\mu \partial_{\t} + \t^{-1} D^{a} V_\mu D_{a},
\ee
from which one finds 
\be \label{delxigamma}
\delta_\xi  \gamma_{\mu \nu}  = -2 V_{(\mu} L_{\nu)}+ 2 D_a V_{(\mu} D^a F_{\nu)}.
\ee

We now consider the hyperboloid components of $\gamma_{\mu \nu}$, defined in \eqref{defsigma}. From \eqref{delxigamma} and the identities presented in section \ref{usefulids} their transformation rules under \eqref{xiapp} (with $R^{\mu \nu}=0$) are found to be
\ba
\delta_\xi \sigma &=& l \label{delxisigma} \\
\delta_\xi \gamma_a &=& l_a - f_a - \partial_a f \\
\delta_\xi \gamma_{ab} &=& 2 D_{(a} f_{b)} + 2 f h_{ab},
\ea
with $f$ and $f_a$ given by \eqref{fitoF}  and similarly for $l$ and $l_a$.  Since  the latter are constructed out of a \emph{constant} vector, they satisfy the additional  properties
\be \label{idsl}
l_a = - \partial_a l, \quad D_a D_b l -  l h_{ab}=0.
\ee

As mentioned in section \ref{timeinfsec}, the gauge invariant content of the gravitational field at this order is captured by the leading electric and magnetic components of the Weyl tensor \cite{hansen,dehouck}
\ba
\E_{ab}  &= &  D_{\langle a} D_{b\rangle}  \sigma \\
\B_{ab}  &= &\frac{1}{2} \epsilon_{a m n} D^{m} k^n_{\phantom{n} b} \label{defB}
\ea
where the angle brackets denote symmetrization and extraction of the trace and 
\be \label{defkab}
k_{ab} = \gamma_{ab} + 2 D_{(a} \gamma_{b)}+ 2 \sigma h_{ab}.
\ee
From \eqref{delxisigma} and \eqref{idsl} one can verify that $\delta_\xi \E_{ab}=0$. On the other hand, it is easy to see that
\be \label{delxik}
\delta_\xi k_{ab} = -2 (D_a D_b f -   h_{ab}f),
\ee
which in turns implies $\delta_\xi \B_{ab}=0$.\\

The leading  $\t \t$ component of Einstein equations constraint $\sigma$ to satisfy \eqref{eqsigma}, while the $\t a$ and $ab$ components  restrict, respectively, the divergence and laplacian of $k_{ab}$ by \emph{homogenous} equations, since the corresponding components of the the matter stress tensor vanish at this order, see \cite{cgw,gianni2}. For spacetimes of physical interest a stronger condition holds on $k_{ab}$, namely the vanishing of the magnetic Weyl curvature \eqref{defB}
\be \label{weylBzero}
\B_{ab} =0,
\ee
which in turn implies \cite{hansen,cgw}
\be \label{kabitoPhi}
k_{ab} = -2 (D_a D_b \Phi -   h_{ab} \Phi),
\ee
where  $\Phi$ is an unconstrained scalar field on the hyperboloid. We will later verify that the asymptotic scattering metric \eqref{Eoneint} indeed  satisfies \eqref{kabitoPhi}.  

From  \eqref{delxik} one finds that $\Phi$ transforms as
\be
\delta_\xi \Phi = f,
\ee
which is why it is identified as a  Goldstone mode for supertranslations \cite{phigoldstone}. 

\subsubsection*{Gauge vs. non-gauge diffeos}

Let us summarize our discussion so far. For spacetimes satisfying \eqref{weylBzero},  the initial ten components in $\gamma_{\mu \nu}$ are reduced to five:\footnote{$\B_{ab}$ is symmetric and traceless and so it has  five independent components.}
\be
\B_{ab} =0 \implies \gamma_{\mu \nu} \leftrightarrow \left\{
\begin{array}{l}
\sigma \\
\gamma_a \\
\Phi
\end{array}
\right. .
\ee

Out of these, we need to isolate ``gauge'' vs ``non-gauge'' components, which requires us to  take into account asymptotic diffeomorphisms (other than asymptotic Lorentz rotations and translation, which do not induce shifts in $\gamma_{\mu \nu}$). These  are parametrized by
\begin{equation} 
\xi^\mu \leftrightarrow 
\left\{
\begin{array}{l}
l \\
f_a \\
f
\end{array}
\right.,
\end{equation}
with $l$ satisfying \eqref{idsl}. We are now faced  with the subtle issue of disentangling pure gauge diffeos vs. asymptotic symmetries. The standard way to discern between the two is by studying the corresponding Hamiltonians \cite{RT}.  We will not go into this type of analysis here, but instead summarize the different prescriptions discussed in the literature.

In all the literature we are aware of, the diffeos $f_a$ are  treated as pure gauge. For instance, in the geometric description of Ashtekar and Hansen \cite{hansen} they do not feature at all, while  in the Beig-Schmidt treatment \cite{BS}, they are fixed by the condition
\be \label{gammaazero}
\gamma_a=0 \implies  f_a=   l_a  - \partial_a f   \quad \text{Beig-Schmidt gauge}.
\ee
Likewise, the harmonic gauge condition fixes them, see \eqref{defga} below.

The treatment of \cite{hansen} does not either feature log translations and only deals with the diffeos spanned by $f$ (referred to as ``Spi supertranslations'' in the spatial infinity context). In \cite{hansen} these are  frozen by imposing the vanishing of $k_{ab}$,
\ba \label{kabzero}
k_{ab}=0 & \implies&  D_a D_b f -   h_{ab}f=0  \\
&\implies& f= -V_\mu T^\mu, \, \text{ with  } \partial_a T^\mu=0.
\ea
The ``residual'' $f$-diffeomorphisms left out by this condition are just spacetime translations, parametrized by constant vectors $T^\mu$.

However,  as argued in \cite{clmassive} (see also \cite{dehouck,phigoldstone,cedric,cgw}) in order to realize BMS supertranslations at timelike infinity, one needs to impose a condition weaker than \eqref{kabzero}. In its more general form, as given  in \cite{dehouck}, this condition fixes the trace of $k_{ab}$ to a given function $\bar k$ on the hyperboloid
\be \label{condf}
k^a_a= \bar k  \implies (D^2 -3)f =0 .
\ee
The family of functions $f$ satisfying \eqref{condf} can then be shown to be in one-to-one correspondence with BMS supertranslations \cite{clmassive}.

From the Beig-Schmidt perspective,  \eqref{condf} appears as an additional condition on top of \eqref{gammaazero}. Interestingly, as reviewed below, in  harmonic gauge this condition appears (almost) automatically.

\subsubsection*{Harmonic gauge residual diffeos} \label{harmggeapp}
Using \eqref{partialmuitohyp}, the harmonic gauge condition \eqref{ggecond} on \eqref{Eone} reads
\be
 V_\nu \ov{1}{e}^{\mu \nu}+ D^a V_\nu D_a \ov{1}{e}^{\mu \nu} =0,
\ee
or, in  terms of the hyperbolic components defined in section \ref{hyperapp}:
\ba
D^b \ov{1}{e}_{ab}-3 \ov{1}{e}_a &=&0, \\
-2 \ov{1}{e}_\parallel-  \ov{1}{e}_\perp + D^b \ov{1}{e}_b &=&0. 
\ea
Applying a general asymptotic diffeo on these equations leads to 
\ba
 g_a &:=& (D^2 +1)f_a + 2 \partial_a f + 2 \partial_a l=0 ,\label{defga}\\
 g &:=&  -2 D^b  f_b  - D^2 f - 3 f  +2 l =0. \label{defg}
\ea
These are the asymptotic harmonic gauge fixing equations. The first one is the analogue of Eq. \eqref{gammaazero}, as it determines $f_a$ in terms of $f$ and $l$ (although this time through a differential  equation).  The second one constrains $f$ and $l$.  To find how,  consider first the divergence of the first equation
\be\label{divg}
D^a g_a = D^2 D^a f_a - D^a f_a + 2 D^2 f + 6 l.
\ee
Solving  \eqref{defg} for $D^b  f_b$ and substituting in \eqref{divg} one finds
\be \label{divgongeqz}
D^a g_a|_{g=0} = -\frac{1}{2}\left(D^2+1\right)\left( (D^2-3) f - 4 l \right) =0, 
\ee
where we used $D^2 l = 3 l$ to factor out the operator $(D^2+1)$. To study \eqref{divgongeqz}, let us rewrite it
as a system of two equations for three unknowns
\ba
\left(D^2+1\right) \psi &=&0  \label{d2plus1psi} \\
 (D^2-3) f &= & 4 l - \psi. \label{d2feqpsi}
\ea

By general arguments of Laplace-type equations in hyperbolic space (see e.g. \cite{solo}) one knows that solutions to \eqref{d2plus1psi} behave as $\psi=O(\ln \rho/\rho)$ for large $\rho$ (see also appendix B.1 of \cite{coito} for explicit integral solutions). On the other hand, Eq. \eqref{d2feqpsi} admits  a formal integral solution for $f$ in terms of the Green's function $\G$ \eqref{Glambda}, with $4l - \psi$ playing the role of source. However, since $l=O(\rho)$, $\G=O(1/\rho^3)$ and the volumen element scales as $\sqrt{h}=O(\rho)$, the integral is logarithmically divergent.\footnote{The coefficient of this divergence is proportional to $\int d^2 \nh (L \cdot n)/(V \cdot n)^3 \sim L \cdot V$.} We thus conclude that $l$ must vanish.

The resulting system of equations still allows for a non-trivial $\psi$. We do not know whether this possibility is  allowed by the full original set of equations, \eqref{defga} and \eqref{defg}. The $\psi=0$ case leads to the   vector fields considered in \cite{clmassive}, which requires  the asymptotic vanishing of their divergence.\footnote{We are excluding superrotations in the comparison with \cite{clmassive}. The analysis of \cite{clmassive} did not consider Eq. \eqref{defga}, but  the present discussion shows that this equation is consistent with the  requirement $\delta_\xi \gamma=0$.} In our notation this  condition reads  $0=\delta_\xi \gamma= 2 D^b f_b + 6 f + 2l$, which, in conjunction to \eqref{defg} leads to $(D^2-3) f = 4 l=0$.

\subsubsection*{Vanishing of the asymptotic magnetic Weyl curvature} 
We now show that the asymptotic metric  \eqref{Eoneint} leads to a tensor $k_{ab}$  of the form \eqref{kabitoPhi}, which in turn implies  $\B_{ab}=0$. We start by rewriting  \eqref{defkab}  in terms of the hyperbolic components of $ \ov{1}{e}_{\mu \nu}$,
\be \label{kabitoemunu}
k_{ab}= 2 \ov{1}{e}_{ab}+ 4 D_{(a} \ov{1}{e}_{b)} -2 \ov{1}{e}_\perp h_{ab} .
\ee
 Using \eqref{Eoneint} and treating separately the trace-free and trace part of \eqref{kabitoemunu} leads to
\ba
k_{\langle ab \rangle} &=& 4 G \int d^3 x' \frac{(\chi^2-3)\partial_{\langle a} \chi \partial_{b\rangle} \chi }{(\chi^2-1)^{3/2}} \rho(x') , \label{kabintegral}\\
k&=&16 G  \int d^3 x' \frac{\chi^2}{\sqrt{\chi^2-1}} \rho(x') .
\ea
From the identities given in appendix \ref{usefulids}, one can show that $ D_{\langle a} D_{b \rangle} f(\chi) = f''(\chi) \partial_{\langle a} \chi \partial_{b\rangle} \chi$ for any function $f(\chi)$. This allow us to pull out the derivatives in \eqref{kabintegral} and write
\be \label{ktfitoPhi}
k_{\langle ab \rangle} = -2 D_{\langle a} D_{b \rangle} \Phi
\ee
with
\be \label{explicitPhi}
\Phi := -2  G  \int d^3 x' \left( \sqrt{\chi^2-1}+\chi \tanh^{-1}(\tfrac{\chi}{\sqrt{\chi^2-1}})\right) \rho(x') .
\ee
Finally, using that $D^2 f(\chi)=f''(\chi) (\chi^2-1)+ 3 f'(\chi) \chi$ one can show that
\be \label{kitoPhi}
k= -2 (D^2-3) \Phi.
\ee
Ecs. \eqref{ktfitoPhi} and \eqref{kitoPhi} imply \eqref{kabitoPhi}, with $\Phi$ given by \eqref{explicitPhi}.

\section{Log soft factor in terms of particle momenta } \label{recoverysahoosenapp}

In this appendix we make contact with the type of expressions presented in \cite{sahoosen,rewritten,senreview}.  Let us start by unpacking \eqref{hlog}. 
The explicit expression of $ h^{(1) \div}_{\mu\nu} $ in terms of the deviation vector is
\be \label{hdev}
 h^{(1) \div}_{\mu\nu} = 4 G \sum_i \left(   \frac{p^i_{\mu}  p^{i}_{\nu} }{p_i \cdot n} c_i \cdot n -p^i_{(\mu} c^{i}_{\nu)}  \right).
\ee
Combining this with the first term in \eqref{hlog} leads to 
\be \label{hlnggeinv}
h^{(\log)}_{\mu\nu}  =  4 G \sum_i    \frac{p^i_{\mu}  p^{i}_{\nu} }{p_i \cdot n} (c_i-c_{\scri^+}) \cdot n -  4 G \sum_i  p^i_{(\mu} c^{i}_{\nu)} .
\ee
Notice that in this form, the invariance under log translations is manifest: The first sum is invariant term by term, while the second sum is invariant thanks to momentum conservation. It is also clear that there is no individual contribution from outgoing massless particles in the first sum since $c^\mu_i=c^\mu_{\scri^+}$ for them. 

For simplicity, we now restrict attention to the case where there is incoming radiation.  The log soft factor then takes the form
\be \label{hlnmassive}
h^{(\log)}_{\mu\nu}  =  4 G \sum_{\substack{i  \\ m_i \neq 0}}    \frac{p^i_{\mu}  p^{i}_{\nu} }{p_i \cdot n} (c_i-c_{\scri^+}) \cdot n -  4 G \sum_i  p^i_{(\mu} c^{i}_{\nu)} \quad \text{(no incoming radiation)}.
\ee
In  harmonic gauge, the deviation vectors in \eqref{hlnmassive} are
\be \label{cmuharm}
\begin{array}{lll}
 c^\mu_{\scri^+} &=& 2 G P^\mu_{\total +} \\
&&\\
c^\mu_\pm &=&  \czero^{\mu}_\pm \pm 2 G P^\mu_{\massless \pm}
 \end{array} \quad \quad \text{(harmonic log frame)}  
\ee
where $\czero^{ \mu}_\pm $ is  given in Eq. \eqref{czeropm} and  $P^\mu_{\massless -}=0$ in the current case of no incoming radiation. Substituting \eqref{cmuharm} in \eqref{hlnmassive}, and splitting the second sum in \eqref{hlnmassive} into massive and massless contributions, one gets
\begin{multline} \label{hlnrewritten}
h^{(\log)}_{\mu\nu}  =  4 G \sum_{\substack{i  \\ m_i \neq 0}} \left(    \frac{p^i_{\mu}  p^{i}_{\nu} }{p_i \cdot n} \czero_i \cdot n -    p^i_{(\mu} \czero^{i}_{\nu)} \right)  \\ +
 8 G^2  n \cdot P_{\massive+} \sum_{\substack{i  \in \out \\ m_i \neq 0}}    \frac{p^i_{\mu}  p^{i}_{\nu} }{p_i \cdot n} +   8 G^2  n \cdot P_{\massive-} \sum_{\substack{i  \in \inn \\ m_i \neq 0}}    \frac{p^i_{\mu}  p^{i}_{\nu} }{p_i \cdot n}   \\
 + 8 G^2   P^{\massive +}_\mu   P^{\massive +}_\nu- 8 G^2   P^{\massive -}_\mu   P^{\massive -}_\nu.
\end{multline}

The first two lines can be written in terms of double sums over massive particles, leading to an expression that has the same form as the full log soft factor, 
except that massless contributions are excluded. The last line then compensates for this exclusion. 
Eq. \eqref{hlnrewritten} corresponds to the Sahoo-Sen ``rewritten" version of the log soft factor \cite{rewritten,senreview}.

It is also interesting to see how  \eqref{hlnrewritten} is recovered by considering \eqref{hlnggeinv} in the future radiative frame. There, $c_{\scri^+}=0$ and  \eqref{hlnggeinv} simplifies to
\be \label{hlnBS}
h^{(\log)}_{\mu\nu}  =  4 G \sum_{\substack{i  \\ m_i \neq 0}}  \left(  \frac{p^i_{\mu}  p^{i}_{\nu} }{p_i \cdot n} c_i \cdot n -    p^i_{(\mu} c^{i}_{\nu)}\right) \quad  \text{($\rad^+$ log frame, no incoming radiation)} .
\ee
The deviation vectors are now given by
\be \label{cmuBS}
c^\mu_\pm =  \czero^{\mu}_\pm \mp 2 G P^\mu_{\massive \pm}  \quad \text{(rad$^+$ log  frame)}  .
\ee
Substituting \eqref{cmuBS} in \eqref{hlnBS} one can check that the expression coincides with  \eqref{hlnrewritten}.

\end{document}